\documentclass[apjl]{emulateapj}
\usepackage{mathptmx}

\begin{document}

\title{Strong Gravitational Lensing and the Structure of Quasar Outflows}
\author{Doron Chelouche\altaffilmark{1}}
\altaffiltext{1} {School of Physics and Astronomy and the Wise
                Observatory, The Raymond and Beverly Sackler Faculty of
                Exact Sciences, Tel-Aviv University, Tel-Aviv 69978,
                Israel; doron@wise.tau.ac.il}
\shortauthors{Chelouche D.}
\shorttitle{Strong Gravitational  Lensing and the Structure of Quasar Outflows}

\begin{abstract}

We show that by analyzing the spectra of lensed broad absorption line quasars (BALQSOs), it is possible to reveal key properties of the outflowing gas in the inner regions of these objects. This results from the fact that each image of the quasar corresponds to a different line of sight through the outflow. This combined with dynamical estimates for the location of the flow, adds new information concerning the lateral, non line of sight structure of the absorbing gas. Here we consider a sample of lensed BALQSOs and note that the similarity of BAL profiles of different images of the same quasar implies that the flow is relatively isotropic on small scales. We show that its geometry is inconsistent with the ballistically accelerated spherical cloud model, and that wind models provide a better description of the flow structure. Furthermore, observations  seem to disagree with naive interpretations of recent time-dependent wind simulations. This may hint on several important physical processes that govern the structure and dynamics of such flows. Future prospects for the study of quasar outflows with the effect of strong gravitational lensing are discussed.
\end{abstract}


\section{Introduction}

The standard model for quasars consists of a central continuum
emitting region (possibly an accretion disk), broad and narrow line
regions (BLR and NLR respectively), and a torus. Among these
components only the NLR can be resolved in nearby
quasars to reveal a stratified structure and a
cone-shape geometry. The physical nature of the other components is
subject to an ongoing debate, e.g., it is not known whether the
BLR is in the form of virialized clouds (e.g., Kaspi \& Netzer 1999)
or a continuously outflowing wind (e.g., Murray, Chiang, Grossman, \&
Voit 1995; hereafter MCGV). 

Observations indicate that
outflowing gas is a common phenomenon in quasars. Such gas is
intrinsic to the quasar, as has been shown for the high velocity
($\gtrsim 10^3~{\rm km~s^{-1}}$), broad absorption line (BAL) flows in
BALQSOs (Weymann 1977). The quasar unification scheme (e.g., Elvis
2000) suggests that BAL flows are ubiquitous and their detection
(in some $15\%$ of type-I objects; Hewett \& Foltz 2003) requires a
fortuitous line of sight to the central engine.

There have been several unsuccessful attempts to constrain the
location of BAL flows via the detection of  variations in velocity or
ionization.  de Kool et al. (2001) have detected
absorption from excited levels of iron which allowed location estimates
in a subclass of BALQSOs (FeLoBALQSOs). An
indirect measure of the gas location relies on the observationally motivated assumption that it is accelerated by radiation pressure force (Arav 1996). This has been the focus of many works concerning the dynamical modeling of such flows. 

Dynamical models consist of cloud (e.g., Chelouche \& Netzer
2001, hereafter CN01), wind (MCGV, Proga, Stone, \& Kallman 2000;
hereafter PSK), and cloud-wind models (e.g., Arav, Li, \& Begelman
1994; hereafter ALB). Some versions of such models can be brought to
some agreement with observations. They do not however agree among
themselves concerning the mass flow rate and other important
parameters. Thus, despite some 30 years of research, several key properties of BAL flows remain unknown. 

In this {\it Letter} we show that strong lensing can add invaluable
information on the structure of BAL flows: In \S 2 we define our
sample of lensed 
BALQSOs. \S 3 discusses the observational constraints on
several flow models. Future prospects and caveats are discussed in \S
4. Summary follows in \S 5.

\section{Lensed BALQSOs: Geometry and Sample}

\begin{figure}
\plotone{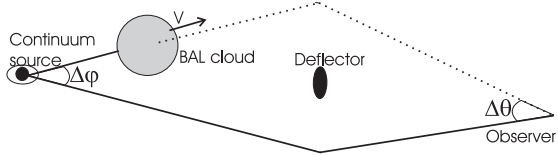}
\caption{A schematic (not to scale) view of a double image lens where each line of sight to the continuum source traverses different sections of the BAL flow (see text).}
\label{lens_geo}
\end{figure}

In gravitationally lensed systems, each image corresponds to a
different line of sight to the central object. For typical lens
configurations (figure \ref{lens_geo}), the angle between the line of
sights, $\Delta 
\varphi$, is of order of the separation angle between the
images ($\Delta \theta$; see e.g., Schneider, Ehlers, \& Falco 2000).  

We consider 7 lensed and one lens candidate (UM\,425; see however
Aldcroft \& Green 2003 who favour a binary quasar interpretation for this
system) BALQSOs whose properties  were taken from
CASTLES\footnote{http://cfa-www.harvard.edu/castles/} (Mu\~{n}oz et
al. 1998) and   
Michalitsianos et al. (1997), and are given in table 1. Assuming a
typical type-I quasar continuum with $\alpha_{ox}=1.4$ (Green et
al. 1995) and a X-ray (0.1-10\,keV) continuum slope,  $\alpha_x=0.9$,
the mean demagnified bolometric luminosity of the quasars in our
sample is $\sim 10^{46}~{\rm erg~s^{-1}}$ (Chartas 2000; this value
depends on the assumed lens model and is somewhat insecure for double
imaged lenses). 

Spectroscopic observations show that in most lensed BALQSOs, the
  absorption troughs are similar between different images to within S/N and
resolution limits. This is probably not a selection effect (due to
the search criteria used for finding lensed objects) since the
fraction of BALQSOs among lensed quasars is larger (Chartas 2000; or
similar but not smaller) than
their fraction among quasars (Hewett \& Foltz 2003). This means that
BAL flows are relatively isotropic on small angular
scales. Nevertheless, for two objects in our sample
(APM\,08279+5255 and the lens candidate UM\,425) there seem to be
noticeable differences in the BAL troughs between the images. We
discuss the implications of these results on various dynamical models
in \S 3. 

\section{Implications for quasar outflow models}

Here we show that by comparing BAL features of different images
one can put constraints on existing flow models. This results from the
different characteristic angular scale of the flow predicted by
different models. Below we discuss each class of models separately. 

\subsection{Single cloud models}

This model was initially invoked to explain the BLR gas and only later
applied to BAL flows. Several 
works (e.g., Blumenthal \& Mathews 1979, hereafter BM79) have demonstrated a qualitative agreement between model and observations in terms of the flow velocity and spectral features. 

In these models the clouds are spherical to a good approximation (BM79) and are confined by some external pressure, $P$ ($P\propto r^{-2}$ is typically assumed where $r$ is the distance from the central object; ALB). The clouds are photoionized by the quasar radiation field and their ionization structure depends primarily on the ionization parameter, $U_x$ (defined as the ratio of photon density in the range 0.1-10\,keV to gas number density; CN01). The terminal velocity of the cloud ($v_{10000}$; in units of $10^4~{\rm km~s^{-1}}$) is given by
\begin{equation}
v_{10000}\simeq 0.1L_{45}^{1/2}r_{17}^{-1/2}M^{1/2},
\label{dynloc}
\end{equation}
where $r_{17}=r/10^{17}$\,cm  is the launching distance, $L_{45}$ the
bolometric luminosity in units of $10^{45}~{\rm erg~s^{-1}}$, and $M$
the ``force multiplier'' (the ratio of the total to the Compton
radiation pressure force; $M(r)=$\,const. for the assumed
pressure profile; CN01).  For $10^{-5}<U_x<0.1$  and the typical velocities and column densities of BAL flows, $M$ can be approximated up to a factor $\sim 3$ by
\begin{equation}
M\simeq 0.3ZU_x^{-1},
\label{m}
\end{equation}
where $Z$ is the gas metalicity normalized to solar values. 

High S/N spectroscopy combined with detailed
photoionization calculations can constrain the ionization
parameter, $U_x$ and the column density $N_{21}$ (in units of
$10^{21}~{\rm cm^{-2}}$; assuming $Z=1$) of the outflowing gas. These seem to cluster around $U_x\sim10^{-4},~N_{21}\gtrsim 1$  for BAL flows (Arav et al. 2001) and $U_x\sim10^{-5},~N_{21}\gtrsim 1$ for FeLoBAL flows (de Kool et al. 2002). Therefore, high ionization BAL flows that are launched from the centers ($r_{17}\gtrsim50$) of luminous quasars ($L_{45}=10$) would reach  a terminal velocity $\lesssim 15,000~{\rm km~s^{-1}}$ (equations \ref{dynloc} \& \ref{m}) which is in qualitative agreement with observations.

Combining equations \ref{dynloc} \& \ref{m} we obtain, 
\begin{equation}
r_{17}\simeq 3\times 10^{-3}v_{10000}^{-2}L_{45}ZU_x^{-1},
\label{loc}
\end{equation}
where $v_{10000}$ is  inferred from the blue wing of BAL troughs. This
estimate for $r_{17}$ can be compared to those derived for
FeLoBALQSOs: For FBQS\,1044+3656 ($L_{45}=10~v_{10000}\simeq 0.5$)
de Kool et al. (2002) conclude that $r_{17}\simeq2\times10^4$ (see
however Everett, K\"{o}nigl, \& Arav 2002) which is consistent with
the dynamical estimate for this object ($r_{17}\sim 10^4$).   
\scriptsize 
\begin{center}
{\sc TABLE 1} \\
{\sc Properties of the Lensed BALQSOs Sample}
\vskip 4pt %
\begin{tabular}{lcccccccc} 
\hline
\hline 
Object & $z_s$ &  $z_l$ & $\Delta \theta$ & $t_{\rm lag}$ & $v_{10000}$ & Refs. \\ 
${\rm [1]}$ & [2] & [3] & [4] & [5] & [6] & [7]  \\
\hline
APM\,08279+5255 & 3.91 & -- & 0.38'' & -  & 1  & a,b\\
SBS\,1520+530 &  1.86 & 0.72 & 1.56'' & 130 & 2  &  c,d\\
HE\,2149-2745 &  2.03 & 0.49 & 1.7'' & 103 & 0.8  & c,e\\
RX J\,0911.4+0551 &  2.8 & 0.77 & 0.48'' & 146 & 0.3 &  c,e\\
PG\,1115+080 & 1.72 & 0.37 & 2.3'' & 25 & 0.6  &  c,f\\
UM\,425 & 1.46 & 0.6 & 6.5'' & -  & 1.2 &  g,h,i\\
H\,1413+117 &  2.55 & -- & 1.35'' & - & 0.9 & c,j\\
J\,1004+1229 &  2.66 & 0.95 & 1.54'' & - & 1  & k\\
\hline
\end{tabular}
\vskip 8pt
\parbox{3.485in}{
\small\baselineskip 9pt 
\footnotesize 
\indent 
The sample of lensed (and the lens candidate UM\,425) BALQSOs: [1]\,Object's name, [2]\,Source redshift, [3]\,Lens redshift, [4]\,Image separation (arcseconds), [5]\,Time lag between images (days), [6]\,Maximum flow velocity ($10000~{\rm km~s^{-1}}$) inferred from the absorption line troughs, [7]\,References: (a)\,Chartas et al. 2002, (b)\,Irwin et al. (1998), (c)\,Chartas (2000), (d)\,Burud et al. (2002), (e)\,Bade et al. (1997), (f)\,Michalitsianos et al. 1996, (g)\,Michalitsianos et al. 1997, (h)\, Green et al. (2001), (i)\, Aldcroft \& Green (2003) ,(j)\,Hutsemekers (1993), (k)\,Lacy et al. (2002)}.
\label{objects}
\end{center}
\normalsize 

It follows from the definition of $U_x$ that the
angle subtended 
by a spherical cloud with a column density
$Z^{-1}N_{21}$ (e.g., Arav et al. 2001) is
\begin{equation}
\Delta \varphi_c\simeq 10 U_x r_{17} L_{45}^{-1} Z^{-1} N_{21}~{\rm arcseconds}.
\label{cloud_omega}
\end{equation}
By combining equations \ref{loc} and \ref{cloud_omega} we obtain 
\begin{equation}
\Delta \varphi_c \simeq  3\times 10^{-2} v_{10000}^{-2}  N_{21}\,{\rm arcseconds}.
\label{cmodel}
\end{equation}
Thus, for all objects in our sample the cloud's lateral size,
$\sim 10^{11}r_{17}\Delta \varphi_c~{\rm cm}$, is smaller than the
Schwartzschild radius ($R_s$) of the central black hole (BH; using the
luminosity--BH mass relation of Kaspi et al. 2000). The fact that BAL
flows only partially cover the ionizing source (whose size is $\sim
10R_s$ for a standard accretion disk; Arav et al. 2001) corroborates
our size estimates.  

Inspection of table 1  reveals that for all objects, $\Delta \varphi_c
\ll \Delta \theta$. For clouds to
show identical absorption features in all quasar images, their aspect
(lateral to radial) ratio should reach $\sim 10^3$ for the more
extreme cases in our sample (the cloud model is however consistent with the
different BAL troughs in e.g., UM\,425; Aldcroft \& Green 2003). Such
oblate clouds would probably be destroyed over dynamical timescales
and a coherent acceleration  is therefore unlikely
(BM79). 

Thus far, we have considered single clouds. Nevertheless, the
flow column density  may be distributed among
many clouds (this is likely given the smooth nature of
BAL troughs). For simplicity we assume that the flow consists of
$n$ identical clouds, each with a column $Z^{-1}N_{21}/n$. In this case,
$\Delta \varphi_c$  is reduced by a factor
$n$. This model is sometimes referred to as the tube model where the
aspect ratio of the flow is much smaller than unity (contrary to what
is implied by lensed quasar observations). The probability for a
perfect alignment of the tube to several lines of sight is negligibly
small . Thus, more likely geometries are those of a sheet or a cone
which are  naturally accounted for by wind models.

\subsection{Clumpy wind models}

In this model (ALB), the flow is assumed to consist of 
numerous small clouds (``cloudlets''). Isotropy is assumed in the
statistical sense, 
i.e.,  each line of sight corresponds to a {\it different} realization
of the {\it same} cloud distribution. Thus, each image of the quasar
may show different absorption features due to statistical fluctuations
in the number of clouds between different lines of sight which follow the
Poissonian distribution. Following ALB, we assume identical
cloudlets. Denoting the total optical depth by $\tau$ and the fluctuations in the optical depth by $\Delta \tau$, one can estimate (assuming 1$\sigma$ fluctuations) the average number of clouds along the lines of sight,
\begin{equation}
n\simeq  \tau ^2 \Delta \tau^{-2}
\label{taus}
\end{equation}
for $\Delta \tau < \tau$.  As the spectral signature of a single cloud cannot
extend considerably beyond one thermal width in velocity space
(typically $\sim 10~{\rm km~s^{-1}}$) it allows us to  constrain, or
put lower limits to the number of clouds in each velocity bin for
non-saturated troughs. In most cases, BAL profiles of different images
are identical within S/N limits implying that the number of clouds in
each velocity bin is $\gg1$ and their total number $\gg 100$.  This
information is orthogonal to that concerning the filling factor of
flows. Furthermore, observing  
systematic differences over a wide velocity range would indicate
changes in the global flow properties (in the cloud
distribution function). Such variations seem to be implied by the
dissimilarities in BAL troughs for two objects in our sample (APM\,08279+5255 and UM\,425; see  \S 4.1) although higher quality data are needed to confirm this point.

\subsection{Continuous wind models}

Time independent continuous wind models usually assume gas which expands
according to the spherical continuity condition $\rho \propto \Omega r^{-2}v^{-1}$ ($\rho, ~v$ being the
gas density and velocity respectively, and $\Omega$ the opening
angle; e.g., MCGV). This assumption
is justified for $\Omega \lesssim 4\pi$. However, flows are likely to evaporate
quickly through their rims for  considerably
smaller $\Omega$, since the expansion timescale ($\propto
\sqrt{\Omega} r_{17}/v_s$, $v_s\simeq 20~{\rm km~s^{-1}}$ is the sound speed)
is smaller than the dynamical timescale ($\propto r_{17}/v_{10000}$). Put
differently, for $\sqrt{\Omega/4\pi}\ll 400v_{10000}^{-1}$\,arcseconds,
the gas would 
evaporate before reaching its terminal velocity. Therefore, a
confining pressure must be invoked, and it is highly unlikely
that the resulting continuity condition would mimic a spherical
expansion.  If the
differences in BAL troughs in APM\,08279+5255 and UM\,425 are real,
these imply that $\sqrt{\Omega/4\pi} \simeq \Delta \theta$ (otherwise
the probability for detecting such
differences would be negligible). Thus, at least for these objects
(see also Chelouche \& Netzer 2003), 
the  assumption of spherically expanding flows is unwarranted  and alternative
geometries with very different spectral predictions  are more
appropriate [e.g., confined
(K\"{o}nigl \& Kartje 1994) and freely expanding flows (Chelouche
2003)].

\subsection{Time dependent wind models}

Time-dependent calculations (PSK) show that Rayleigh-Taylor,
Kelvin-Helmholtz, and radiative instabilities result in a filamentary
sub-structure of the flow. Currently, there are no spectral
predictions for their model and direct comparison with observations is
complicated. Nevertheless, the results for the angular distribution of
$U_x$ and density (see figure 4a,b in PSK for the angular range
subtended by high velocity gas) imply that the extreme ($>10$ orders
of magnitude!) variations over $\lesssim 10\degr$  result from
opacity rather than density effects. Thus, $U_x$ is a measure of the
transmitted spectrum. Assuming that $U_x$ is a continuous function of angle
down to small scales, then interpolation of PSK results
predicts it should differ by many orders of magnitude over $\sim$1
arcsecond scales.  Such extreme variations would result in 
different spectral signatures between quasar images which are not
observed. A more detailed study of this effect (e.g., by accounting
for the finite size of the continuum source and the stratified
flow structure) is limited by the
low resolution of current simulations.  This apparent contradiction
between model and observations may be resolved if there is a typical
scale (larger than $r\Delta \theta \sim 10^{14}\,$cm) below
which variations are suppressed (e.g., due to the dissipation of
turbulence below some scale or by the stablizing effect of magnetic
fields; e.g., Proga 2003). 

\section{Future prospects}

We have demonstrated a new method for
investigating the structure of BAL flows. Similar reasoning can be
applied to additional types of flows namely, narrow absorption line
(NAL) and X-ray BAL (XBAL) flows. For NAL flows $N_{21}\sim2$
  (e.g., Kraemer et al. 2002; but
see also  Hamann, Netzer, \& Shields
2000 for lower values) and
$v_{10000}\lesssim 0.1$ (Crenshaw et al. 1999) and, therefore, $\Delta \varphi_c\gtrsim 2$ arcseconds,
i.e., the single cloud model cannot be firmly ruled out for typical
lens separations. This stems from the fact that such gas  lies at
larger distances and is more dilute. XBALs have so far been observed in
 two objects (APM\,08279+5155 and PG\,1115+080; Chartas et
al. 2002, 2003) and seem to have $N_{21}=10^2$, and
$v_{10000}=10$. This gives $\Delta \varphi_c\sim10^{-2}$
arcsecond. Nevertheless, due to the complications associated
with  X-ray spectroscopy, it is difficult to compare
 XBALs of different images.

Given several lines of sight, it may be possible to detect whether the
flow is collimated and at what direction. From pure
geometrical considerations, a collimated flow will show a velocity
difference $\delta v\simeq v \delta \varphi$ between two lines of
sight $\delta \varphi$ apart. For SBS\,1520+530, $\delta v \sim
0.1~{\rm km~s^{-1}}$, i.e., within the resolution of next generation
spectrometers. This, combined
with an independent measure for the orientation of the quasar
(e.g., via the detection of a jet), may allow us to ascertain whether
BAL flows are predominantly polar or planar and how they 
relate to  other components of the quasar.

Outflows are likely to rotate around the central object (e.g., MCGV)
and it may be possible to detect spectral absorption features sweeping across our line of sight over Keplerian timescales,
\begin{equation}
\tau_{\rm Kep.}\simeq  10^{-6}v_{10000}^{-3}L_{45}Z^{3/2}U_x^{-3/2}\Gamma^{1/2}\Delta \varphi\,{\rm days},
\label{kep}
\end{equation}
($\Gamma$ is the luminosity in units of the Eddington
luminosity). Typically  $\tau_{\rm Kep.}\sim 100$\,days for BAL flows
in quasars with $\Gamma=0.1$. Detecting such events would set more
stringent limits on the geometry of the outflowing gas and may reveal
the angular momentum balance in the inner quasar regions (perhaps even
in the accretion disk). Such studies may also provide better estimates
for $H_0$ since absorption events do not depend on the global geometry of the system and are less smeared in time compared to emission events.


\subsection{Caveats}

We have assumed that flow properties do not
change on timescales shorter than the time lag, $t_{\rm lag}$, between
quasar images. This requires that both
the dynamical timescale, $t_{\rm dyn}$ and the
intrinsic flux variability timescale, $t_{\rm flux}$, are longer than $t_{\rm
  lag}$. Indeed, the dynamical timescale, $t_{\rm dyn} \simeq
v_{10000}^{-3}L_{45}ZU_x^{-1}$\,days is much longer than $t_{\rm lag}$
(table 1) for UV flows ($t_{\rm dyn}\ll t_{\rm lag}$ for XBALs) and recent
variability studies of high redshift  quasars indicate that  $t_{\rm
  flux}\gg t_{\rm lag}$ (Kaspi et al. 2003). Thus, any differences
between UV BAL troughs are probably not related to time-lag effects.  

An additional complication arises from micro-lensing (ML) by stars in
the lensing galaxy which can differentially (de-)magnify the continuum
emitting region (e.g., Wambsganss 1992). Its effect, being 
  line of sight dependent (e.g., Ofek \& Maoz 2003), may alter the
  shape of non-black troughs and
  result in apparent discrepancies between them. A more
  quantitative discussion requires further study (Chelouche 2003, in
  preparation) and is beyond the scope of this {\it   Letter}. 


\section{Conclusions and Summary}

In this {\it Letter} we show that by analyzing the spectral absorption
features of spatially resolved lensed BALQSOs, one can obtain
invaluable information concerning the geometry and sub-structure of
BAL flows. Given the currently available spectroscopic data we
conclude that such  flows are relatively isotropic on a few arcsecond
scales. Our analysis suggests that the coherently accelerated cloud
model (and the related tube model) are inconsistent with observations,
and that BAL flows are probably made of one or several outflowing
sheets or cones with lateral dimensions much larger than the typical
separation between quasar images.  The similarity of the absorption
features between different images is inconsistent with naive
interpretations of time-dependent dynamical simulations. This may
indicate  a typical length scale below which 
sub-structure is suppressed.  Future high resolution spectroscopy of
lensed BALQSOs may reveal whether such flows are collimated and
distinguish between polar and planar flow configurations. Multi-epoch   
observations may put further constraints on the lateral geometry and
dynamics of BAL flows as well as lead to better estimates for
$H_0$. Such information cannot be obtained from observations of
non-lensed BALQSOs, and is imperative to our understanding of quasar
structure and the physical processes near black-holes. 

\acknowledgments

The author wishes to thank H. Netzer, I. Goldman, O. Shemmer, and
E. O. Ofek for many helpful discussions, and the anonymous referee for
his valuable comments. The Dan David Prize scholarship is gratefully acknowledged.

\end{document}